\documentclass[12pt]{article}
\usepackage{epsfig}
\usepackage{latexsym}
\usepackage{amssymb}
\usepackage{amsfonts}
\newcommand{\be}{\begin{equation}}
\newcommand{\ee}{\end{equation}}
\newcommand{\bea}{\begin{eqnarray}}
\newcommand{\eea}{\end{eqnarray}}

\newcommand{\ep}{\varepsilon}
\newcommand{\sg}{\sigma}

\newcommand{\pa}{\partial}
\newcommand{\dd}{\mbox{d}}

\newcommand{\nn}{\nonumber}

\newcommand{\SA}{\left\{ \begin{array}{ll}}
\newcommand{\Sa}{\left[ \begin{array}{ll}}
\newcommand{\FA}{\end{array}\right.}

\newcommand{\leftdot}{$\!\!\!\mbox{\bf{.}}\,\,$}
\newtheorem{tr}{Theorem}
\newcommand{\BTh}{\begin{tr}\leftdot}
\newcommand{\ETh}{\end{tr}}
\newtheorem{lmm}{Lemma}
\newcommand{\BLm}{\begin{lmm}\leftdot}
\newcommand{\ELm}{\end{lmm}}
\newtheorem{deff}{Definition}
\newcommand{\BDf}{\begin{deff}\leftdot}
\newcommand{\EDf}{\end{deff}}
\newtheorem{stttt}{Step}
\newcommand{\BStt}{\begin{stttt}\leftdot}
\newcommand{\EStt}{\end{stttt}}
\newenvironment{proof}{$\blacktriangleleft$ }{$\blacktriangleright$
\indent}
\newcommand{\BPr}{\begin{proof}}
\newcommand{\EPr}{\end{proof}}
\newcommand{\BZM}{\begin{zamet}}
\newcommand{\EZM}{\end{zamet}}

\begin{document}
\parindent=1.5pc

\begin{titlepage}
\rightline{February 2006}

\bigskip
\begin{center}
{{\large\bf
An Algorithm to Construct Gr\"obner Bases for Solving
Integration by Parts Relations} \\
\vglue 5pt
\vglue 1.0cm
{\large  A.V. Smirnov}\footnote{E-mail: asmirnov@rdm.ru
\\
Supported in part by RFBR, grant 05-01-00988
}\\
\baselineskip=14pt
\vspace{2mm}
{\normalsize 
Scientific Research Computing Center of
Moscow State University}
\baselineskip=14pt
\vspace{2mm}
\vglue 0.8cm
{Abstract}}
\end{center} 
\vglue 0.3cm
{\rightskip=3pc
 \leftskip=3pc
\noindent
This paper is a detailed description of an algorithm
based on a generalized Buchberger algorithm for
constructing Gr\"obner-type bases associated with polynomials of
shift operators. The algorithm is used for calculating
Feynman integrals and has proven itself efficient in several complicated
cases.
\vglue 0.8cm}
\end{titlepage}

\section{Introduction}

Feynman integrals over loop momenta are building blocks of  quantum-theoretical
amplitudes in the framework of perturbation theory.
After a tensor reduction based on some projectors a
given Feynman graph generates various scalar Feynman integrals
that have the same structure of the integrand with various
distributions of powers of propagators which we shall also call
{\em indices}:
\bea
  F(a_1,\ldots,a_n) &=&
  \int \cdots \int \frac{\dd^d k_1\ldots \dd^d k_h}
  {E_1^{a_1}\ldots E_n^{a_n}}\,.
  \label{eqbn}
\eea
Here $k_i$, $i=1,\ldots,h$, are loop momenta
and the denominators $E_r$ are either quadratic or linear with respect
to the loop momenta $p_i=k_i, \; i=1,\ldots,h$, or
independent external momenta $p_{h+1}=q_1,\ldots,p_{h+N}=q_N$ of the graph.
(Linear propagators appear naturally in effective field theories
and in asymptotic expansions of Feynman integrals in various
limits.)

Irreducible polynomials in the numerator can be represented as
denominators raised to negative powers.
Usual prescriptions $k^2=k^2+i 0$, etc. are implied.
The dimensional regularization \cite{dimreg} with $d=4-2\ep$  is assumed.
The Feynman integrals are functions of the masses, $m_i$,
and kinematic invariants, $q_i\cdot q_j$. However, we shall
omit this dependence because we shall pay special attention to
the dependence on the indices. We shall also omit the dependence
on $d$.


At the modern level of perturbative calculations, when one needs
to evaluate millions of Feynman integrals
(\ref{eqbn}), a well-known
optimal strategy here is to derive, without calculation, and then
apply some relations between the given family of Feynman integrals
as {\it recurrence relations}.
A well-known standard way to obtain such relations is provided
by the method of integration by parts
(IBP) \cite{IBP}.
Practically, one starts from IBP relations
\bea
\int\ldots\int \dd^d k_1 \dd^d k_2\ldots
\frac{\pa}{\pa k_i}\left( p_j
\frac{1}{E_1^{a_1}\ldots E_n^{a_n}}
\right)   =0  \;.
\label{RR}
\eea
After the differentiation, resulting scalar products,
$k_i\cdot k_j$ and $k_i\cdot q_j$
are expressed in terms of the factors in the denominator,
and one arrives at
IBP relations which can be written as 
%
\begin{equation}
\sum c_i F(a_1+b_{i,1},\ldots,a_n+b_{i,n})
=0\,,
\label{IBP}
\end{equation}
where $b_{i,j}$ are integer,
$c_i$ are polynomials in $a_j$,
$d$, masses $m_i$ and kinematic invariants,
and $F(a_1,\ldots,a_n)$ are Feynman integrals (\ref{eqbn}) of the
given family.

One tries to use IBP relations in order to express a general dimensionally
regularized integral of the
given family as a linear combination of some
`irreducible' integrals which
are also called {\it master} integrals.
There are several recent attempts to make this reduction procedure
systematic, in particular, the so-called Laporta's algorithm
\cite{LGR1,LGR2} (There is a public version of implementing the
corresponding algorithm on a computer \cite{AnLa}.) and Baikov's
method (see \cite{Bai,ST} and chapter 6 of \cite{EFI}). Another activity in this direction is connected
with the use of Gr\"obner bases \cite{Buch}. The first variant of
this approach was suggested in \cite{Tar1}, where IBP relations
were reduced to differential equations. First attempts to use
directly the non-commutative Gr\"obner bases
in the algebra generated by shift operators were made in
\cite{Gerdt,GeRo}.
In the previous paper \cite{ourwork}
(see also \cite{rc05} for a brief review) we presented another approach
based on Gr\"obner bases. Its specifics lies in using more information
on the given family of Feynman integrals, in particular the
boundary conditions, i.e. the conditions of the following form:
\be
F(a_1,a_2,\ldots,a_n)=0\mbox{ when }a_{i_1}<0,\ldots a_{i_k}<0
\label{boundary}
\ee
for some set of indices $i_j$ (for example, we always have
$F(a_1,a_2,\ldots,a_n)=0$ if all $a_i$ are non-positive).
Since that time the algorithm has been made more efficient
by introducing the so-called s-form. Moreover, the algorithm
has been made about ten times faster.
Here it will be described in details (the
paper \cite{ourwork} was designed to give an introduction to the
method and not all definitions were formal).

\section{Preliminaries}

To describe the algorithm we have to introduce some notation.
Let $\mathcal K$ be the field of rational functions of physical
variables $m_i$, $q_i\cdot q_j$, $d$, and $\mathcal{A}$
be the algebra\footnote{An algebra
over a field is a vector space over this field and a ring at the
same time.} over $\mathcal K$ generated by elements $Y_i$,
$Y^-_i$
and $A_i$ with the following relations:
\bea
Y_i Y_j &=& Y_j Y_i,\;\;\;\ A_i A_j = A_j A_i, \;\;\;Y_i A_j = A_j
Y_i+\delta_{i,j}Y_i,
\\ \nn
Y^-_i Y^-_j &=& Y^-_j Y^-_i,\;\;\;
Y^-_i Y_j = Y_j Y^-_i, \;\;\;
Y^-_i A_j = A_j Y^-_i-\delta_{i,j}Y_i,\;\;\;Y^-_i Y_i=1\,
\eea
where $\delta_{i,j}=1$ if $i=j$ and $0$ otherwise.
For convenience we will write $(Y^-_i)^k=Y^{-k}_i$.
Let $\mathcal F$ be the field of functions of $n$ integer
arguments $a_1,a_2,\ldots,a_n$.
The algebra $\mathcal{A}$
acts on this field\footnote {
({\em i}) for any $a\in\mathcal{A}$ and $f\in \cal F$
we have an element $a\cdot f\in\cal F$;
({\em ii}) for any $a,b\in\mathcal{A}$ and $f,g\in \cal F$ we have
$(a+b)\cdot(f+g)=a\cdot f+a\cdot g+b\cdot f+b\cdot g$;
({\em iii}) for any $a,b\in\mathcal{A}$ and $f\in \cal F$ we have
$(ab)\cdot f=a\cdot(b\cdot f)$.}, where
\bea
(Y_i\cdot F)(a_1,a_2,\ldots,a_n)=F(a_1,\ldots,a_{i-1},a_i+
1,a_{i+1},\ldots,a_n)\,,
\\\nn
(A_i \cdot F)(a_1,a_2,\ldots,a_n)=a_i F(a_1,a_2,\ldots,a_n)\,.
\eea


The left-hand sides of relations (\ref{IBP}) can be represented
as elements of the algebra
${\mathcal{A}}$ applied to $F$; we will denote these elements by $f_1,\ldots,f_k$.
Now, for $F(a_1,\ldots,a_n)$ defined by (\ref{eqbn}), we have
\be
f_i\cdot F=0\mbox{ or }(f_i\cdot F)(a_1,\ldots,a_n)=0
\ee
for all $i$. Let us
generate a (left) ideal $\cal I$ by the elements $f_1,\ldots,f_k$.
We will call $\cal I$ the \textit{ideal of the IBP relations}.
Obviously,
\be
f\cdot F=0\,, \;\; \mbox{or} \;\;
(f\cdot F)(a_1,\ldots,a_n)=0 \;\; \mbox{ for any }\;\;f\in {\cal I}\,.
\ee

Let us consider the set $\cal D$ with elements $\{c_1,c_2,\ldots,c_n\}$
where all $c_i$ are equal to $1$ or $-1$. The elements of this set
will be called \textit{directions}. For any direction
$\nu=\{c_1,\ldots,c_n\}$ we will consider a region
$\sg_\nu = \{ (a_1,\ldots,a_n): (a_i - 1/2)  c_i > 0\}$
and call it {\em a sector}.
Obviously the union of all sectors contains all the integer points
in the $n$-dimensional vector space and the intersection of any two
sectors is an empty set.
For a sector $\sg_\nu$ we will say that its direction is $\nu$.


We will say that an element $X\in \mathcal{A}$ is written in
the {\em proper} form if it is represented as
\be
X=\sum r_j(A_1,\ldots,A_n)\prod_i Y^{d_{i,j}}_i,
\ee
where $r_j$ are polynomials (with coefficients in $\mathcal K$) and $d_{i,j}$ are integers.
(So, all the operators $A_i$ are placed on the left from
the operators $Y_i$.)
Obviously any element $X\in\mathcal{A}$ has a unique proper
form. We will say that an element of $\mathcal{A}$
is a \textit{monomial} if in its proper form only one coefficient function $r_j$
is non-zero.

Let $\mathbb N^n=\{(b_1,\ldots,b_n)\}$ where all $b_i$ are integers
and $b_i\geq 0$.
This is a semi-group
(with respect to $(b_1,\ldots,b_n)+(b'_1,\ldots,b'_n)=(b_1+b'_1,\ldots,b_n+b'_n)$).
We will say that an ordering on $\mathbb N^n$ is
\textit{proper} if
\\
i) for any $a\in \mathbb N^n$ not equal to $(0,\ldots 0)$ one has $a\succ (0,\ldots 0)$
\\
ii) for any $a,b,c\in \mathbb N^n$ one has $a\succ b$
if and only if $a+c\succ b+c$.

Let us fix a direction $\nu=\{c_1,c_2,\ldots,c_n\}$
We will say that the \textit{$\nu$-degree} of a monomial
$r(A_1,\ldots,A_n)\prod_i Y^{d_{i}}_i$
is $\{d_1 c_1,\ldots,d_n c_n\}$ if all the products $d_i c_i$ are
non-negative
and undefined otherwise.

Let us choose a proper ordering.
Take an element $X\in\mathcal{A}$ and write it in its proper form.
Consider the set of $\nu$-degrees of all the monomials in the
decomposition of $X$. If this set is empty (all the degrees
are undefined) we will say that the $\nu$-highest degree
of $X$ is undefined as well, otherwise we will say that
$\nu$-highest degree of $X$ is the highest element of that
set in terms of the fixed ordering.

We will say that a direction $\{c_1,\ldots c_n\}$ is lower than
$\{c'_1,\ldots c'_n\}$ if $c_1\leq c'_1,\ldots,c_n\leq c'_n$
and they are not equal. The
same is said about the corresponding sectors. We will say that a
sector $\sg$ is \textit{trivial} if all the integrals
$F(a_1,\ldots,a_n)$ are zero for $(a_1,\ldots,a_n)\in\sg$
due to boundary conditions (\ref{boundary}). The same will be said
about the direction of the sector.

\section{The s-reduction algorithm}

Suppose that for each non-trivial sector $\sg_\nu$ we are given
a finite basis $\{X_{\nu,1},\ldots,X_{\nu,k_\nu}\}\subset\cal I$
and an ordering.

We are going to describe an algorithm called s-reduction.

Input: a linear combination of integrals $F(a_1,\ldots,a_n)$.

Output: another linear combination of integrals containing
the integrals that could not be reduced by these
bases (an integral is called $F(a_1,\ldots,a_n)$ \textit{irreducible}
for the given sets of orderings and bases if the
$\nu$-reduction of $F(a_1,\ldots,a_n)$ returns
$F(a_1,\ldots,a_n)$, where $\nu$ is such a direction
that $(a_1,\ldots,a_n)\in\sg_\nu$).

If one chooses properly the orderings and the elements $X$ then
there is a finite number of irreducible integrals,
therefore the s-reduction will always output a combination of
them but such a choice is a non-trivial procedure.
The algorithm
that might construct such elements will be described in the next section.
But first we have to explain what s-reduction is.
Some examples of the bases can be found at
http://www.srcc.msu.ru/nivc/about/lab/lab4\_2/index\_eng.htm

\textit{S-Reduction}
\begin{tabbing}
1. \= $L$=Input; $M=0$.
\\
2. \> \textbf{While} \= $L\neq 0$
\\
\>\> 3. \= Let $S$ be the set of sectors that contain a point
$(a_1,\ldots,a_n)$ where
\\\>\>\> $F(a_1,\ldots,a_n)$ has a non-zero
coefficient in the decomposition of $L$.
\\
\>\> 4. \> Let $\nu$ be a direction such that $\sg_\nu\in S$ and there is
no other sector
\\\>\>\> $\sg_{\nu'}\in S$ such that $\sg_\nu$ is lower
than $\sg_{\nu'}$.
\\\>\> 5. \> Let $L=L_1+L_2$ where $L_1$ contains those and only those
\\ \>\>\> $F(a_1,\ldots,a_n)$ where $(a_1,\ldots,a_n)\in\sg_\nu$.
\\ \>\> 6. \> $N$ = $\nu$-reduction of $L_1$ by $\{X_{\nu,1},\ldots,X_{\nu,k_\nu}\}$ (to be described below)
\\ \>\> 7. \> Let $N=N_1+N_2$ where $N_1$ contains those and only those
\\ \>\>\> $F(a_1,\ldots,a_n)$ where $(a_1,\ldots,a_n)\in\sg_\nu$.
\\ \>\> 8. $L=L_2+N_2$; $M=M+N_1$.
\\ 9. \> Output=$M$
\end{tabbing}

In a few words we are using the $\nu$-reduction of $X$ starting
from higher sectors. Basically,
it is a formalization of the standard method
people use to solve the IBP relations ``by hand''.
 Now we are going to describe what
$\nu$-reduction is.

\textit{$\nu$-reduction}

Input: direction $\nu$; an element $L$ that is a linear combination of integrals
$F(a_1,\ldots,a_n)$ such that all
$(a_1,\ldots,a_n)\in\sg_\nu$ ; a finite basis $\{X_1,\ldots,X_k\}\subset\cal I$.

Output: another linear combination of integrals $F(a_1,\ldots,a_n)$ such that
$(a_1,\ldots,a_n)\in\sg_{\nu'}$ where $\nu'=\nu$ or $\nu'$ is lower
than $\nu$.

\begin{tabbing}
1. $\;$\= Let $p=\{p_1,\ldots,p_n\}=\{(c_1+1)/2,\ldots,(c_n+1)/2\}$, where
$\nu=\{c_1,\ldots,c_n\}$.
\\ 2. \> Set $Y=0$.
\\ 3. \> Let $X\in\mathcal{A}$ be the element obtained by replacing
$F(a_1,\ldots,a_n)$
\\ \> with $\Pi_i\; Y_i^{a_i-p_i}$ in $L$. One has $L=(X\cdot F)(p)$.
\\ 4. $\;$ \> \textbf{Whi}\=\textbf{le} $X\neq 0$
\\ \>\> 5. $\;$ \=Let $U$ be the highest term of $X$, $U=C\; \Pi_i Y_i^{d_i c_i}$.
\\ \>\> 6. \> For all possible products $T=(\Pi_i Y_i^{x_i}) \cdot X_j$ such
that the proper form of $T$
\\ \>\>\> has a non-zero coefficient at $\Pi_i Y_i^{d_i c_i}$
\\ \>\>\> \textbf{Do} \= (there is a finite number of possibilities)
\\ \>\>\>\> 7. $\;$\= Replace all $A_i$ in the proper form of $T$ with $p_i$.
\\ \>\>\>\> 8. \> If for some $j$ such that $c_j=-1$ there is a term of $T$ with $Y_j$
\\ \>\>\>\>\> in a positive degree then continue the cycle with the next element.
\\ \>\>\>\> 9. \> If the coefficient $C'$ of $T$ at $\Pi_i Y_i^{d_i c_i}$ is equal to zero
\\ \>\>\>\>\> then continue the cycle with the next element.
\\ \>\>\>\> 10. \> Take $Z=X-(C/C')T$. Let $Z_1$ be the $\nu$-sector part of $Z$
\\ \>\>\>\>\> (the sum of monomials in the decomposition of $Z$
\\ \>\>\>\>\> that have a defined degree) and $Z_2=Z-Z_1$.
\\ \>\>\>\> 11. \> If the $\nu$-highest degree of $Z_1$ is lower than the
\\ \>\>\>\>\> $\nu$-highest degree of $X$ then replace $X$ with $Z_1$,
\\ \>\>\>\>\> $Y$ with $Y+Z_2$ and go to the start of the While cycle (step 4).
\\ \>\> 12. \> Replace $X$ with $X-U$ and $Y$ with $Y+U$
\\ 13. \> Return $(Y\cdot F)(p)$
\end{tabbing}

The idea of the algorithm is to represent the given linear
combination as an element $X$ of $\mathcal{A}$ being applied to $F$ and the
value being taken in the corner of the sector. Then one tries all
possible transformation of the obtained element that lower the
$\nu$-degree of $X$ (this is a generalization of the standard
reduction procedure). The point is that through all the algorithm
the value $((X+Y)\cdot F)(p)$ is not changed. This is based on the
fact that $(Z\cdot F)(p)=0$ for any $Z\in\cal I$.
Note that the elements $X$ and $Y$ in their proper forms
do not contain the operators $A_i$ (because of the replacement in step 7).
Therefore in step 9 we have $C'\in\mathcal K$ and
$C'\neq 0$, so the division is possible.
The step 12 is the
place where the terms that cannot be reduced are moved from $X$ to
$Y$. If the basis is chosen properly, those terms will correspond
to a finite number of master integrals.

\section{Constructing s-bases}

Let us fix a non-trivial direction $\nu=\{c_1,c_2,\ldots,c_n\}$
and let $p=\{p_1,\ldots,p_n\}=\{(c_1+1)/2,\ldots,(c_n+1)/2\}$.
Our task is to construct an \textit{s-basis} for this direction,
i.e. such a basis $\{X_{1},\ldots,X_{k}\}\subset\cal I$ that the
$\nu$-reduction for this sector has only a finite number of
irreducible integrals.

The relations (\ref{IBP}) give us a basis of $\cal I$ but
generally it is not an s-basis. Moreover, one has to choose an
appropriate ordering.

We are going to describe an algorithm that takes the relations
(\ref{IBP}) as input and aims to construct an s-basis.
The results of the algorithm greatly depend on the
choice of the ordering and in complicated cases the proper choice
seems to be unique. Now suppose we have fixed an ordering and an
initial basis $\{X_{1},\ldots,X_{k}\}\subset\cal I$. Let us
describe our algorithm.


First of all let us define the \textit{s-form} of an element
$X\in \mathcal{A}$ (note that this definition and some more definitions
below depend on the choice of the direction $\nu$, but it is fixed
in this section). So, the \textit{s-form} of an element
$X\in \mathcal{A}$ is an element $T$ of the form $(\Pi_i Y_i^{x_i}) \cdot X$
satisfying the following properties:
\\
(\textit{i}) The $\nu$-highest degree of $T$ is defined and for any integer $(y_1,\ldots,y_n)$ such that
$y_1 c_1\geq 0,\ldots,y_n c_n\geq 0$ the $\nu$-highest degree
of $(\Pi_i Y_i^{y_i}) \cdot T$ is equal to the $\nu$-highest degree of $T$
plus $(y_1,\ldots,y_n)$;
\\
(\textit{ii}) The $\nu$-highest coefficient of $(\Pi_i Y_i^{y_i}) \cdot T$ does not
vanish when $A_1=p_1,\ldots,A_n=p_n$ for
all integer $(y_1,\ldots,y_n)$ such that
$y_1 c_1\geq 0,\ldots,y_n c_n\geq 0$;
\\
(\textit{iii}) For all $j$ such that $c_j=-1$ the degrees of $Y_i$ in the
proper form of $T$ are non-positive;
\\
(\textit{iv}) The numbers $(c_1 x_1,\ldots,c_n x_n)$ are minimal possible for
all $(x_1,\ldots,x_n)$ satisfying the properties (\textit{i})-(\textit{iii}).

Let us reformulate this definition less formally. The properties (\textit{i}), (\textit{ii})
mean that this element has ``enough'' terms whose degrees lie in
the sector $\sg_\nu$. It is needed so that this element can be used for the
$\nu$-reduction. The property (\textit{iii}) means that if $c_j=1$
then both $Y_j$ and $Y^-_j$ can appear in the s-form, but if $c_j=-1$
then only $Y^-_j$ can. The meaning of this requirement is
to control that using this element
for $\nu$-reduction we will not obtain elements that lie in
sectors higher than $\nu$.

Through all the algorithm we will store a basis of $\cal I$
consisting of elements in the s-form. Let us describe how the
reduction of an element of $\mathcal{A}$ via a basis
$\{X_{1},\ldots,X_{k}\}\subset\cal I$ works.

\begin{tabbing}
1. \= $Y$=s-form of Input.
\\ 2.\>If $Y=0$ then Return 0.
\\ 3.\>For all $j$ such that all the numbers $d_i-d'_i$ are positive,
where $(d_1,\ldots,d_n)$ is
\\\> the $\nu$-highest degree of $Y$ and $(d'_1,\ldots,d'_n)$ is the $\nu$-highest degree of $X_j$
\\\> \textbf{Do} \=
\\\>\> 4. \= Let $Z=C' \;Y - C\; \Pi_i Y_i^{c_i(d_i-d'_i)} \cdot X_j$,
where $C$ is the $\nu$-highest coefficient
\\ \>\>\> of $Y$ and $C'$ is the $\nu$-highest coefficient of
$(\Pi_i Y_i^{c_i(d_i-d'_i)}) \cdot X_j$.
\\ \>\> 5. \> $Z$ = s-form of $Z$.
\\ \>\> 6. \> If the $\nu$-highest degree of $Z$ is lower than
$\nu$-highest degree of $Y$ then
\\ \>\>\> replace $Y$ with $Z$ and Go to step 2.
\\
8. \> Return $Y$.
\end{tabbing}

It is easy to see that this reduction stops after a finite number
of steps (there can't be an infinite sequence of decreasing
degrees). Basically, this procedure is close to the standard
reduction procedure in the Buchberger algorithm. The
difference is the usage of s-forms and the fact that the elements $Y_j$ can have
both positive and negative degrees.

Now we can describe the main algorithm. As it has been said
earlier, it starts from a basis of $\cal I$, moreover, all
elements are taken in their s-forms. The goal of the algorithm is
to construct another basis such that the
$\nu$-reduction for this sector has only a finite number of
unreducible integrals. Therefore, after an element is added to the
basis or a basis element is replaced we are performing a test to
verify this condition (Completion Criteria). It consists of
checking, whether for any $j$ there is such $m$ that
for all integer $l\geq m$ the element $F(p_1,\ldots,p_{j-1},p_j+l c_j,p_{j+1},\ldots,p_n)$
can be reduced via this basis in the $\nu$-reduction algorithm.
\\ Here is the main algorithm:
\begin{tabbing}
1. \=\textbf{While} \= not Completion Criteria
\\\>\> 2. \=If there is an element in the basis that can be reduced by some
other \\\>\>\> element,
replace it with the result of the reduction and
restart the cycle.
\\
\>\> 3. \>Choose a pair of elements of the basis $X'$ and $X''$.
\\
\>\> 4. \>Choose the smallest
possible integers $(d_1,\ldots,d_n)$ such that
$d_j\geq d'_j$ and \\\>\>\> $d_j\geq d''_j$,
where
$(d'_1,\ldots,d'_n)$ is the $\nu$-highest degree of $X'$ and
$(d''_1,\ldots,d''_n)$ \\\>\>\> is the $\nu$-highest degree of $X''$.
\\
\>\> 5. \>Evaluate
$Z=C''\;\Pi_i Y_i^{c_i(d_i-d'_i)}\cdot X'  - C' \; \Pi_i Y_i^{c_i(d_i-d''_i)}\cdot X''$,
where $C'$ is the
\\\>\>\> $\nu$-highest coefficient of
$(\Pi_i Y_i^{c_i(d_i-d'_i)})\cdot X'$
and $C''$ is the $\nu$-highest
\\\>\>\>coefficient of $(\Pi_i Y_i^{c_i(d_i-d''_i)})\cdot X''$,
--- the s-polynomial of $X'$ and $X''$.
\\
\>\> 6. \>$Z$ = Reduction of $Z$.
\\
\>\> 7. \>If $Z\neq 0$ then add $Z$ to the basis.
\end{tabbing}

When implementing the present algorithm it is natural to store
the information about the pairs
where the s-polynomials have been evaluated to avoid
repeating the same calculations.
Of course, the choice performed at line 3 might be different,
and the algorithm effectiveness greatly depends on this choice.
One more improvement of the
algorithm is the use of symmetries of the diagram. Instead of
evaluating an s-polynomial one might take an element symmetric to
some element of the basis and reduce it the same way.
All this is realized by introducing a function on the set
of pairs of elements of the basis and another one on the set
of all possible symmetric element to the elements of the basis.
So, at line 3 we are choosing such an element to evaluate,
that the value of this \textit{choice function} is minimal.
Currently the algorithm uses the choice function intended
to minimize the degree of the resulting element,
and it already makes the algorithm effective.
The work on finding the optimal choice functions is in progress.

\section{Conclusion}

The algorithm described above is close to the Buchberger
algorithm, but has a significant difference. The main problem is
that generally one can have two elements of the same degree that
cannot be reduced one by another in this algorithm.
Therefore, the standard proof that the algorithm has to stop at
some point 
does not work here.
To prove that this algorithm stops for any given family of Feynman
integrals is an open problem which, hopefully, will be solved in
the nearest future.
On the other hand, the present algorithm
appears to be much more efficient
to solve reduction problems for Feynman integrals than more or less
straightforward generalization of the Buchberger algorithm (see, e.g.,  \cite{Gerdt}).
This was demonstrated in \cite{ourwork} where
the reduction problem was solved for Feynman integrals relevant
to the two- and three-loop static quark potential, with the number of
indices $n=7$. (In two loops, well-known results
\cite{PSKPSS} were reproduced.)
New results obtained with the help of this algorithm, in cases
with the number of indices $n=9$, will be published soon \cite{GSS}.
Preliminary analysis shows that the algorithm can work
successfully in problems with the number of indices up to $n=12$.
The extension to more complicated problems will, of course,
require its modifications, in various places.

An implementation of the s-reduction part of the algorithm in
Mathematica can be found at
http://www.srcc.msu.ru/nivc/about/lab/lab4\_2/index\_eng.htm
together with pedagogical examples and at least the s-bases that
have been constructed for the problems considered in \cite{ourwork}.

{\em Acknowledgments.}
I am grateful to V.~Smirnov for having given me the idea
to apply the theory of Gr\"obner bases to evaluating
Feynman integrals, for constant testing of my algorithm
and for permanent support in my work.
I am grateful to A.~Grozin, A. Onishchenko and  M.~Steinhauser for reading my paper and
giving useful advice and comments.

\end{document}